\documentclass[a4paper, 12pt, oneside]{article}
\usepackage[cp1251]{inputenc}
\usepackage[english, russian]{babel}
\usepackage{ graphics,amsfonts, amsmath,bm,amssymb, array}
\usepackage[dvips]{graphicx}
\usepackage[flushleft,small]{caption2}
\makeatletter

\renewcommand{\Re}{\mathop{\rm Re\,}}
\renewcommand{\Im}{\mathop{\rm Im\,}}

\makeatother {

\begin{document}
\thispagestyle{empty} \large
\renewcommand{\abstractname}{\, }
\renewcommand{\refname}{\begin{center}
Список литературы\end{center}}

 \begin{center}
\bf
Analytical solution of second Stokes problem about behaviour of gas over
fluctuating surface by means of ellipsoidal statistical equation
\end{center}%\medskip
\begin{center}
  \bf
A. V. Latyshev\footnote{$avlatyshev@mail.ru$}
and  A. A. Yushkanov\footnote{$yushkanov@inbox.ru$}
\end{center}\medskip

\begin{center}
{\it Faculty of Physics and Mathematics,\\ Moscow State Regional
University, 105005,\\ Moscow, Radio str., 10--A}
\end{center}\medskip

\tableofcontents
\setcounter{secnumdepth}{4}

\begin{abstract}

Second Stokes problem about behaviour of rarefied gas filling half-space
is analytically solved.
A plane limiting half-space makes harmonious fluctuations
in the plane. The kinetic equation with  modelling integral
collisions in form of ellipsoidal statistical model is used .
The case of diffusion reflexions of gas molecules from a wall is considered.
Function distribution of gas molecules is constructed
and  mass velocity of gas also in half-space is found.
Hydrodynamic character of the solution  at small frequencies
of fluctuation plane limiting gas is revealed.
The force of a friction operating from gas on border making in the plane
oscillatory movement is found.


Аналитически решена вторая задача Стокса о поведении разреженного газа,
заполняющего полупространство. Плоскость,
ограничивающая полупространство, совершает гармонические колебания
в своей плоскости. Используется кинетическое уравнение с
модельным интегралом
столкновений в форме эллипсоидально--статистической модели. Рассматривается
случай диффузного отражения молекул газа от стенки. Построена функция
распределения газовых молекул и найдена массовая скорость газа в
полупространстве.
Выявлен гидродинамический характер решения при малых частотах колебания
ограничивающей газ плоскости.
Найдена сила трения, действующая со стороны газа на границу,
совершающую в своей плоскости колебательное движение.

{\it Ключевые слова:} постановка задачи Стокса, разделение переменных,
собственные решения, непрерывный и дискретный спектр, точное решение.
\end{abstract}

\begin{center}
{\bf Введение}
\end{center}

Задача о поведении газа над движущейся поверхностью в последние годы
привлекает пристальное внимание \cite{Ai}--\cite{15}. Это связано с
развитием современных технологий, в частности, технологий наноразмеров.
В \cite{Ai}--\cite{15} эта задача решалась численными или
приближенными методами.

Впервые задача о поведении сплошной среды над стенкой,
колеблющейся в своей плоскости,
была рассмотрена Дж. Г. Стоксом \cite{Stokes}. Сейчас такую задачу
называют второй задачей Стокса \cite{Ai}--\cite{SS-2002}.

В последние годы на тему этой задачи появился ряд публикаций.
В работе \cite{SK-2007} получены коэффициенты вязкостного и теплового
скольжения с использованием различных модельных уравнений. Использованы
как максвелловские граничные условия, так и граничные условия
Черчиньяни---Лэмпис.

В статье \cite{10} рассматривается газовый поток над бесконечной пластиной,
совершающей гармонические колебания в собственной плоскости. Найдена скорость
газа над поверхностью и сила, действующая на поверхность со стороны газа.
Для случая низких частот задача решена на основе уравнения Навье---Стокса.
Для произвольных скоростей колебаний поверхности задача решена численными
методами на основе кинетического уравнения Больцмана с интегралом
столкновений в форме БГК (Бхатнагар, Гросс, Крук).

%Работа \cite{11} является экспериментальным исследованием. Изучается поток
%газа, создаваемый механическим резонатором при различных частотах колебания
%резонатора. Эксперименты показывают, что при низких частотах колебаний
%резонатора, действующая на него со стороны газа сила трения прямо
%пропорциональна частоте колебания резонатора. При высоких частотах
%колебания резонатора ($~10^8$ Гц) действующая на него сила трения от частоты
%колебаний не зависит.

В статье \cite{12} рассматривается пример практического применения
колебательной системы, подобной рассматриваемой во второй задаче Стокса,
в области нанотехнологий.

В диссертации \cite{15} были предложены два решения второй задачи Стокса,
учитывающие весь возможный диапазон коэффициента аккомодации тангенциального
импульса. Эти решения отвечают соответственно гидродинамическому и
кинетическому описанию поведения газа над колеблющейся поверхностью в
режиме со скольжением.

В наших работах \cite{ALY-1}--\cite{ALY-3} дается аналитическое решение
второй задачи Стокса  с использованием БГК (Бхатнагар, Гросс, Крук)
кинетического уравнения. В этих работах отыскиваются собственные функции и соответствующие собственные значения,
отвечающие как дискретному, так и непрерывному спектрам. Исследована структура
дискретного и непрерывного спектров. Развивается математический аппарат,
необходимый для аналитического решения задачи и приложений.
Наконец, в \cite{ALY-3} приводится аналитическое решение.

В настоящей работе строится аналитическое решение второй задачи Стокса
с использованием линеаризованного эллипсоидально статистического
кинетического уравнения. На основе аналитического решения
построена функция распределения газовых молекул
в полупространстве и непосредственно у колеблющейся
границы. %\clearpage

\begin{center}
\item{}\section{Постановка задачи}
\end{center}

Пусть разреженный одноатомный газ занимает полупространство $x>0$
над плоской твердой поверхностью, лежащей в плоскости $x=0$.
Поверхность $(y,z)$ совершает гармонические колебания вдоль оси $y$
по закону $u_s(t)=u_0e^{-i\omega t}$.

Линеаризуем функцию распределения, полагая $f=f_0(1+\varphi)$.
%$f(x,t,\mathbf{v})=f_0(v)(1+\varphi(x,t,\mathbf{v}))$.
Здесь
$f_0(v)=n(\beta/\pi)^{3/2}\exp(-\beta v^2)$ -- абсолютный максвеллиан,
$\beta=m/(2kT)$.
Рассмотрим линеаризованное эллипсоидально--статистическое
кинетическое уравнение (кратко: ЭС--уравнение)
$$
\dfrac{\partial \varphi}{\partial t}+
v_x\dfrac{\partial \varphi}{\partial
x}+\nu\varphi(x,t,\mathbf{v})=$$$$=
2\nu \beta v_yu_y(x,t)+2a\dfrac{\nu \beta^2}{\rho}v_xv_y\sigma_{xy}(x,t).
\eqno{(1.1)}
$$

В (1.1) $\nu=1/\tau$ -- частота столкновений газовых молекул, $\tau$ --
время между двумя последовательными столкновениями молекул, $m$ -- масса
молекулы, $k$ -- постоянная Больцмана, $T$ -- температура газа, $u_y(x,t)$ --
массовая скорость газа, $\sigma_{xy}(x,t)$ -- компонента тензора
вязких напряжений,
$a$ -- числовой параметр уравнения, причем при $a=-1$ число
Прандтля является истинным $(\Pr=2/3)$,
$$
u_y(x,t)=\dfrac{1}{n}\int v_yf(x,t,\mathbf{v})d^3v,
$$
$$
\sigma_{xy}(x,t)=m\int v_xv_y f(x,t,\mathbf{v})d^3v,
$$
$n$ -- числовая плотность (концентрация) газа. Концентрация газа и
его температура считаются постоянными в линеаризованной постановке задачи.

Введем безразмерные скорости и параметры: безразмерную скорость молекул:
$\mathbf{C}=\sqrt{\beta}\mathbf{v}$ \;$(\beta=m/(2kT))$, безразмерную
массовую скорость $U_y(x,t)=\sqrt{\beta}u_y(x,t)$, безразмерное время
$t_1=\nu t$ и
безразмерную скорость колебаний пластины $U_s(t)=U_0e^{-i\omega t}$,
безразмерную компоненту тензора вязких напряжений
$P_{xy}(x,t)=(\beta/\rho)\sigma_{xy}(x,t)$,
где $U_0=\sqrt{\beta}u_0$ -- безразмерная амплитуда скорости колебаний
границы полупространства. Тогда уравнение (1.1) может быть записано в виде:
$$
\dfrac{\partial \varphi}{\partial t_1}+
C_x\dfrac{\partial \varphi}{\partial
x_1}+\varphi(x_1,t_1,\mathbf{C})=$$$$=
{2C_y}U_y(x_1,t_1)+2aC_xC_yP_{xy}(x_1,t_1).
\eqno{(1.2)}
$$

Заметим, что для безразмерного времени
$U_s(t_1)=U_0e^{-i\omega_1t_1}$.

В задаче о колебаниях газа требуется найти функцию распределения
$f(x_1,t_1,\mathbf{C})$ газовых молекул.

Затем на основании найденной функции распределения требуется найти массовую
скорость газа, значение
массовой скорости газа непосредственно у стенки. Кроме того, требуется
вычислить силу сопротивления газа, действующую на колеблющуюся пластину,
ограничивающую газ.
Подчеркнем, что задача о колебаниях газа решается в линеаризованной постановке.
Линеаризация задачи проведена по безразмерной массовой скорости
$U_y(x_1,t_1)$ при условии, что $|U_y(x,t_1)|\ll 1$. Это неравенство
эквивалентно неравенству
$$
|u_y(x_1,t_1)|\ll v_T,
$$
где $v_T=1/\sqrt{\beta}$ -- тепловая скорость молекул, имеющая порядок
скорости звука.

Величины безразмерных массовой скорости и компоненты тензора вязких
напряжений через функцию $\varphi$ выражаются следующим образом:
$$
U_y(x_1,t_1)=\dfrac{1}{\pi^{3/2}}\int \exp(-C^2)C_y\varphi(x_1,t_1,
\mathbf{C})d^3C,
\eqno{(1.3)}
$$
и $$
P_{xy}(x_1,t_1)=\dfrac{1}{\pi^{3/2}}\int \exp(-C^2)C_xC_y\varphi(x_1,t_1,
\mathbf{C})d^3C.
\eqno{(1.4)}
$$

С помощью (1.3) и (1.4) кинетическое линеаризованное уравнение (1.2)
записывается в виде:
$$
\dfrac{\partial \varphi}{\partial t_1}+
C_x\dfrac{\partial \varphi}{\partial x_1}+\varphi(x_1,t_1,\mathbf{C}) =
\dfrac{2C_y}{\pi^{3/2}}
\int\exp(-{C'}^2)C_y'\varphi(x_1,t_1,\mathbf{C'})\,d^3C'+
$$
$$
+\dfrac{2aC_xC_y}{\pi^{3/2}}
\int\exp(-{C'}^2)C_x'C_y'\varphi(x_1,t_1,\mathbf{C'})\,d^3C'
\eqno{(1.5)}
$$

Сформулируем диффузные граничные условия, записанные относительно функции
$\varphi(x_1,t_1,\mathbf{C})$:
$$
\varphi(0,t_1,\mathbf{C})=2C_yU_s(t_1)e^{-i\omega_1t_1},\quad C_x>0,
\eqno{(1.6)}
$$
%и
$$
\varphi(x_1\to+\infty,t_1,\mathbf{C})=0.
\eqno{(1.7)}
$$

Итак, граничная задача о колебаниях газа сформулирована полностью и состоит
в решении уравнения (1.5) с граничными условиями (1.6) и (1.7).

Учитывая, что колебания пластины рассматриваются вдоль оси $y$, будем искать
функцию $\varphi$ в виде
$$
\varphi(x_1,t_1,\mathbf{C})=C_ye^{-i\omega_1t_1}h(x_1,C_x).
\eqno{(1.8)}
$$
С помощью (1.8) получаем следующую граничную задачу:
$$
\mu\dfrac{\partial h}{\partial x_1}+z_0h(x_1,\mu)
=\dfrac{1}{\sqrt{\pi}}
\int\limits_{-\infty}^{\infty}\exp(-{\mu'}^2)(1+a\mu\mu')h(x_1,\mu')d\mu',
\eqno{(1.9)}
$$
$$
h(0,\mu)=2U_0,\qquad \mu>0, \qquad   z_0=1-i\omega_1,
\eqno{(1.10)}
$$
$$
h(+\infty,\mu)=0.
\eqno{(1.11)}
$$

\begin{center}
\item{}\section{Собственные решения непрерывного спектра}
\end{center}
Разделение переменных в уравнении (1.9) осуществляется
следующей подстановкой
$$
h_\eta(x_1,\mu)=\exp\Big(-\dfrac{x_1z_0}{\eta}\Big)\Phi(\eta,\mu),
\eqno{(2.1)}
$$
где $\eta$ -- параметр разделения, или спектральный параметр, вообще говоря,
комплексный.

Подставляя (2.1) в уравнение (1.9) получаем характеристическое уравнение
$$
z_0(\eta-\mu)\Phi(\eta,\mu)=\dfrac{1}{\sqrt{\pi}}\eta n_0(\eta)+
\dfrac{1}{\sqrt{\pi}}a\mu\eta n_1(\eta),
\eqno{(2.2)}
$$
где
$$
n_k(\eta)=\int\limits_{-\infty}^{\infty}
\exp(-{\mu'}^2)\mu'^k\Phi(\eta,\mu')d\mu',\quad k=0,1.
$$

Из уравнения (2.2) находим, что
$$
n_1(\eta)=-\dfrac{i \omega_1}{z_0}\eta n_0(\eta).
$$

Следовательно,
уравнение (2.2) можно представить в виде:
$$
(\eta-\mu)\Phi(\eta,\mu)=\dfrac{1}{\sqrt{\pi}z_0}\eta n_0(\eta)(1-b \mu\eta),
\eqno{(2.3)}
$$
где
$$
b=\dfrac{i\omega_1a}{z_0}.
$$

Далее  примем следующую нормировку
$$
n_0(\eta)\equiv \int\limits_{-\infty}^{\infty}
\exp(-{\mu'}^2)\Phi(\eta,\mu')d\mu'\equiv {z_0}.
$$
Тогда  уравнение (2.3) имеет при
$\eta,\mu\in(-\infty,+\infty)$ решение   \cite{Zharinov}
$$
\Phi(\eta,\mu)=\dfrac{1}{\sqrt{\pi}}P\dfrac{\eta(1-b\mu\eta)}{\eta-\mu}+
e^{\eta^2}\lambda(\eta)\delta(\eta-\mu),
\eqno{(2.4)}
$$
где $\delta(x)$ -- дельта--функция Дирака, символ $Px^{-1}$
означает главное значение интеграла при интегрировании $x^{-1}$,
$\lambda(z)$ -- дисперсионная функция, введенная равенством
$$
\lambda(z)=1-i\omega_1+\dfrac{z}{\sqrt{\pi}}\int\limits_{-\infty}^{\infty}
\dfrac{e^{-\tau^2}(1-b z\tau)d\tau}{\tau-z}.
$$
Эту функцию можно преобразовать к виду:
$$
\lambda(z)=-i\omega_1+(1-b z^2)\lambda_0(z),
$$
где $\lambda_0(z)$ -- известная функция из теории плазмы,
$$
\lambda_0(z)=\dfrac{1}{\sqrt{\pi}}\int\limits_{-\infty}^{\infty}
\dfrac{e^{-\tau^2}\tau d\tau}{\tau-z}.
$$

Собственные функции (2.4) называются собственными функциями непрерывного
спектра, ибо спектральный параметр $\eta$ непрерывным образом заполняет
всю действительную прямую.

Таким образом, собственные решения уравнения (1.9) имеют вид (2.1),
в котором функция $\Phi(\eta,\mu)$ определяется равенством (2.4).

По условию задачи мы ищем решение, невозрастающее вдали от стенки.
В связи с этим спектром граничной задачи будем называть положительную
действительную полуось параметра $\eta$.

Приведем формулы Сохоцкого сверху и снизу на действительной оси
для дисперсионной функции:
$$
\lambda^{\pm}(\mu)=\pm i\sqrt{\pi}\mu e^{-\mu^2}(1-b\mu^2)-i\omega_1+
\dfrac{1-b\mu^2}{\sqrt{\pi}}\int\limits_{-\infty}^{\infty}
\dfrac{e^{-\tau^2}\tau d\tau}{\tau-\mu}.
$$
Разность граничных значений сверху и снизу на действительной оси
дисперсионной функции отсюда равна:
$$
\lambda^+(\mu)-\lambda^-(\mu)=2\sqrt{\pi}\mu e^{-\mu^2}(1-b\mu^2)i,
$$
полусумма граничных значений равна:
$$
\dfrac{\lambda^+(\mu)+\lambda^-(\mu)}{2}=-i\omega_1+\dfrac{1-b\mu^2}{\sqrt{\pi}}
\int\limits_{-\infty}^{\infty}\dfrac{e^{-\tau^2}\tau d\tau}{\tau-\mu}.
$$
Сингулярный интеграл в этих равенствах понимается в смысле главного
значения.

\begin{center}
\item{}\section{Структура дискретного спектра}
\end{center}

Покажем, что дискретный спектр, состоящий из нулей дисперсионного
уравнения $\lambda(z)=0$, содержит два нуля $-\eta_0$ и $\eta_0$,
из которых обозначается через $\eta_0$ тот нуль, у которого $\Re \eta_0>0$.

Сначала рассмотрим случай малых значений $\omega_1$.
Разложим дисперсионную функцию в асимптотический ряд по
отрицательным степеням
переменного $z$ в окрестности бесконечно удаленной точки:
$$
\lambda(z)=-i\omega_1+\dfrac{b}{2}-\dfrac{1}{2z^2}+\dfrac{3b}{4z^2}+
\dots, \quad z\to \infty.
\eqno{(3.1)}
$$

Из разложения (3.1) видно, что при малых значениях $\omega_1$ %и $b$
дисперсионная функция имеет два отличающиеся лишь знаками
комп\-лек\-сно--значных нуля:
$$
\pm\eta_0^{(0)}(\omega_1)=
\sqrt{i\dfrac{1-3i\omega_1a/2z_0}{\omega_1(2-a/z_0)}}=
\sqrt{i\dfrac{1-i\omega_1-3i\omega_1a/2}
{2\omega_1(1-i\omega_1-a/2)}},\quad -1\leqslant a\leqslant 0.
$$

Отсюда видно, что при $\omega_1\to 0$ оба нуля дисперсионной функции
имеют пределом одну бесконечно удаленную точку $\eta_i=\infty$ кратности
(порядка) два.

Теперь исследуем случай произвольных значений $\omega_1$.
Далее нам понадобится функция
$$
G(\tau)=\dfrac{\lambda^+(\tau)}{\lambda^-(\tau)}=
\dfrac{-i\omega_1+(1-b\tau^2)\lambda_0^+(\tau)}
{-i\omega_1+(1-b\tau^2)\lambda_0^-(\tau)}.
\eqno{(3.2)}
$$

Выделим у функции $G(\tau)$ действительную и мнимую части.
Заметим, что
$$
b=b_1+ib_2,\qquad b_1=-\dfrac{a\omega_1^2}{1+\omega_1^2},
\qquad b_2=\dfrac{a\omega_1}{1+\omega_1^2},
$$
$$
\lambda_0^{\pm}(\tau)=l(\tau)\pm is(\tau),\quad
s(\tau)= \sqrt{\pi}\tau e^{-\tau^2},
$$
$$
l(\tau)=1-2\tau^2 \int\limits_{0}^{1}e^{-\tau^2(1-x^2)}dx.
$$

Теперь равенство (3.2) записывается следующим образом: $$
G(\tau)=
\dfrac{-i\omega_1+[(1-b_1\tau^2)-ib_2\tau^2](l(\tau)+is(\tau))}
{-i\omega_1+[(1-b_1\tau^2)-ib_2\tau^2](l(\tau)-is(\tau))},
$$
или
$$
G(\tau)=\dfrac{p+q-i(\omega_1-p_1+q_1)}{p-q-i(\omega_1+p_1+q_1)},
$$
где $$
p(\tau)=(1-b_1\tau^2)l(\tau), \qquad
q(\tau)=b_2\tau^2s(\tau),
$$
$$
p_1(\tau)=(1-b_1\tau^2)s(\tau),\qquad
q_1(\tau)=b_2\tau^2l(\tau).
$$

Теперь функцию $G(\tau)$ можно представить в виде
$$
G(\tau)=G_1(\tau)+iG_2(\tau),
$$
где
$$
G_1(\tau)=\dfrac{g_1(\tau)}{g_0(\tau)},\qquad
G_2(\tau)=\dfrac{g_2(\tau)}{g_0(\tau)},
$$
$$
g_1(\tau)=p^2-q^2+\omega_1^2-p_1^2+q_1^2,
$$
$$
g_2(\tau)=2[pp_1+q(\omega_1+q_1)],
$$
$$
g_0(\tau)=(p-q)^2+(\omega_1+p_1+q_1)^2.
$$
Функции $g_j(\tau) (j=0,1,2)$ понадобятся в явном виде:
$$
g_1(\tau)=\omega_1^2-[s^2(\tau)-l^2(\tau)][(1-b_1\tau^2)^2+b_2^2\tau^4],
$$
$$
g_2(\tau)=2s(\tau)\{\omega_1b_2\tau^2+l(\tau)[(1-b_1\tau^2)^2+b_2^2\tau^4]\},
$$
$$
g_0(\tau)=\omega_1^2+2\omega_1[(1-b_1\tau^2)s(\tau)+b_2\tau^2l(\tau)]+
[l^2(\tau)+s^2(\tau)][(1-b_1\tau^2)^2+b_2^2\tau^4].
$$
В этих равенствах
$$
(1-b_1\tau^2)^2+b_2^2\tau^4=\dfrac{1+\omega_1^2(1+a\tau^2)^2}{1+\omega_1^2}.
$$
Таким образом,  окончательно получаем:
$$
g_1(\tau)-\dfrac{\omega_1^4-\omega_1^2s_1(\tau)-s_0(\tau)}{1+\omega_1^2},
$$
где
$$ s_0(\tau)=s^2(\tau)-l^2(\tau),\qquad s_1(\tau)=s_0(\tau)(1+a\tau^2)^2-1,
$$
и
$$
g_2(\tau)=\dfrac{2s(\tau)}{1+\omega_1^2}\Big\{a\omega_1^2\tau^2+
l(\tau)[1+\omega_1^2(1+a\tau^2)^2]\Big\},
$$
$$
g_0(\tau)=\omega_1^2+[l^2(\tau)+s^2(\tau)]\dfrac{1+\omega_1^2(1+a\tau^2)^2}
{1+\omega_1^2}+$$$$+2\omega_1\Big[\Big(1+\dfrac{a\omega_1\tau^2}{1+\omega_1^2}\Big)
s(\tau)+\dfrac{a\omega_1\tau^2}{1+\omega_1^2}l(\tau)\Big].
$$

Можно показать с помощью принципа аргумента аналогично тому,
как это сделано в \cite{ALY-1}, что число
нулей дисперсионной функции равно:
$$
N=\dfrac{1}{2\pi i}\int\limits_{-\infty}^{\infty}d\ln G(\tau)=
\dfrac{1}{\pi i}\int\limits_{0}^{\infty}d\ln G(\tau)=$$$$=
\dfrac{1}{\pi}\Big[\arg G(\tau)\Big]_{0}^{+\infty}=\dfrac{1}{\pi}\arg G(+\infty)
=2\varkappa(G),
$$
т.е. удвоенному индексу функции $G(\tau)$.

Введем угол $\theta(\tau)=\arg G(\tau)$ -- главное значение аргумента,
фиксированное в нуле условием $\theta(0)=0$,
$$
\theta(\tau)=\arcctg \dfrac{\Re G(\tau)}{\Im G(\tau)}=
\arcctg\dfrac{g_1(\tau)}{g_2(\tau)}.
\eqno{(3.3)}
$$
Из уравнения $g_1(\tau)=0$ найдем его положительный корень:
$$
\omega_1(a)=\sqrt{\dfrac{s_1(\tau)}{2}+
\sqrt{\Big(\dfrac{s_1(\tau)}{2}\Big)^2+
s_0(\tau)}}\equiv\Omega(\tau,a).
$$

Введем выделенную частоту колебаний пластины, ограничивающей газ:
$$
\omega_1^*(a)=\max\limits_{0<\tau<\infty}\Omega(\tau,a).
\eqno{(3.4)}
$$
Эту частоту колебаний будем называть {\it критической}.

Аналогично \cite{ALY-1} можно показать, что в случае, когда частота
колебаний пластины
меньше критической, т.е. при $0\leqslant \omega <\omega_1^*(a)$, индекс
функции $G(t)$ равен единице. Это означает, что число комплексных нулей
дисперсионной функции в плоскости с разрезом вдоль действительной оси,
равно двум.

В случае, когда частота колебаний пластины превышает критическую
($\omega>\omega_1^*(a)$) индекс функции $G(t)$ равен нулю: $\varkappa(G)=0$.
Это означает, что дисперсионная функция не имеет нулей в верхней и нижней
полуплоскостях. В этом случае дискретных (частных) решений исходное
уравнение (1.9) не имеет.

Таким образом, дискретный спектр характеристического уравнения,
состоящий из нулей дисперсионной функции, в случае
$0\leqslant \omega_1<\omega_1^*(a)$ есть множество из двух точек
$\eta_0$ и $-\eta_0$. При $\omega_1>\omega_1^*(a)$ дискретный
спектр --- это пустое множество. При $0\leqslant \omega_1<\omega_1^*(a)$
убывающее собственное решение уравнения (1.9) имеет вид
$
h_{\eta_0}(x_1,\mu)=e^{-x_1z_0/\eta_0}\Phi(\eta_0,\mu),
$ где
$$
\Phi(\eta_0,\mu)=\dfrac{1}{\sqrt{\pi}}\dfrac{\eta_0(1-b\mu\eta_0)}{\eta_0-\mu}
$$
-- собственная функция характеристического уравнения.
Это означает, что дискретный спектр рассматриваемой граничной задачи
состоит из одной точки $\eta_0$ в случае
$0 <\omega_1<\omega_1^*(a)$.
При $\omega_1\to 0$ оба нуля $\pm \eta_0$, как уже указывалось выше,
перемещаются в одну
и ту же бесконечно удаленную точку. Это значит, что в случае $\omega_1=0$
дискретный спектр задачи состоит из одной бесконечно
удаленной точки кратности два
и является присоединенным к непрерывному спектру.
В этом случае дискретных (частных) решения ровно два:
$$
h_1(x_1,\mu)=1, \qquad h_2(x_1,\mu)=x_1-\dfrac{2}{2-a}\mu.
$$

Нетрудно показать, что параметр уравнения $a$ и число Прандтля связаны
соотношением:
$$
\Pr=\dfrac{2}{2-a},\quad \text{откуда}\quad a=-\dfrac{2(1-\Pr)}{\Pr}.
$$

Правильному (истинному) числу Прантля $\Pr=2/3$ отвечает значение параметра
$a=-1$.

Приведем таблицу критических частот в зависимости от значений  числа Прандля и
параметра уравнения $a$ согласно (3.4).

{\bf Таблица} значений критических частот.\medskip
\bigskip

\begin{tabular}{|c|c|c|}
  \hline
Число Прандтля $\Pr$ & Параметр $a$&Критическая частота $\omega_1^*$\\\hline
  1 & 0 & 0.733 \\\hline
  0.952 & -0.1 & 0.717 \\\hline
  0.909 & -0.2 & 0.717 \\\hline
  0.870 & -0.3 & 0.691 \\\hline
  0,833 & -0.4 & 0.681 \\\hline
  0.800 & -0.5 & 0.672 \\\hline
  0.769 & -0.6 & 0.662 \\\hline
  0.741 & -0.7 & 0.654 \\\hline
  0.714 & -0.8 & 0.648 \\\hline
  0.690 & -0.9 & 0.642 \\\hline
  2/3 & -1 & 0.637     \\\hline
\end{tabular}

\begin{center}
\item{}\section{Аналитическое решение граничной задачи}
\end{center}

Составим общее решение уравнения (1.9) в виде суммы частного (дискретного)
решения, убывающего вдали от стенки, и интеграла по
непрерывному спектру от собственных решений, отвечающих непрерывному спектру:
$$
h(x_1,\mu)=a_0\Phi(\eta_0,\mu)e^{-x_1z_0/\eta_0}+
e^{-x_1z_0/\eta}\Phi(\eta,\mu)a(\eta)d\eta.
\eqno{(4.1)}
$$

Здесь $a_0$ -- неизвестный постоянный коэффициент, называемый коэффициентом
дискретного спектра, причем при $\varkappa=0$ этот коэффициент равен нулю:
$a_0=0$, $a(\eta)$ -- неизвестная функция, называемая коэффициентом
непрерывного спектра, $\Phi(\eta,\mu)$ -- собственная функция
характеристического уравнения,
отвечающая непрерывному спектру и единичной нормировке.
Постоянная $a_0$ и функция $a(\eta)$ подлежат нахождению из
граничного условия (1.10).

Разложение (4.1) можно представить в явном  виде:
$$
h(x_1,\mu)=a_0\dfrac{\eta_0(1-b\mu\eta_0)}{\sqrt{\pi}(\eta_0-\mu)}
\exp\Big(-\dfrac{x_1z_0}{\eta_0}\Big)+\hspace{3cm}$$$$+
\exp\Big(-\dfrac{x_1z_0}{\mu}+\mu^2\Big)\lambda(\mu)a(\mu)\theta_+(\mu)+
$$
$$
\hspace{1cm}+\dfrac{1}{\sqrt{\pi}}\int\limits_{0}^{\infty}
\exp\Big(-\dfrac{x_1z_0}{\eta}\Big)\dfrac{\eta(1-b\mu\eta)
a(\eta)d\eta}{\eta-\mu},
\eqno{(4.2)}
$$
где $\theta_+(\mu)$ -- функция Хэвисайда,
$$
\theta_+(\mu)=\left\{\begin{array}{c}
                       1,\qquad \mu>0, \\
                       0,\qquad \mu<0.
                     \end{array}\right.
$$

Очевидно, что разложение (4.2)
автоматически удовлетворяет граничному условию (1.11) вдали
от стенки. Подставим разложение (4.2) в граничное условие (1.10).
Получаем одностороннее сингулярное интегральное уравнение с ядром Коши
$$
\dfrac{\eta_0(1-b\mu\eta_0)a_0}{\sqrt{\pi}(\eta_0-\mu)}+\dfrac{1}{\sqrt{\pi}}
\int\limits_{0}^{\infty}
\dfrac{\eta a(\eta)(1-b\mu\eta)d\eta}{\eta-\mu}+ $$$$+
\exp(\mu^2)\lambda(\mu)a(\mu)=2U_0,\; \qquad\mu>0.
\eqno{(4.3)}
$$

Сингулярный интеграл из (4.3) представим в виде суммы двух слагаемых и
перепишем уравнение в виде:
$$
\dfrac{\eta_0(1-b\mu\eta_0)a_0}{\sqrt{\pi}(\eta_0-\mu)}
+A+\dfrac{1}{\sqrt{\pi}}\int\limits_{0}^{\infty}
\dfrac{\eta(1-b\eta^2)a(\eta)}{\eta-\mu}d\eta+
$$
$$
+ \exp(\mu^2)\lambda(\mu)a(\mu)\theta_+(\mu)=2U_0,\;\qquad \mu>0.
\eqno{(4.4)}
$$
Здесь
$$
A=b \dfrac{1}{\sqrt{\pi}}\int\limits_{0}^{\infty}\eta^2a(\eta)d\eta.
\eqno{(4.5)}
$$

Введем вспомогательную функцию
$$
N(z)= \dfrac{1}{\sqrt{\pi}}\int\limits_{0}^{\infty}
\dfrac{\eta(1-b\eta^2)a(\eta)d\eta}
{\eta-z},
\eqno{(4.6)}
$$
для которой согласно формулам Сохоцкого
$$
N^+(\mu)-N^-(\mu)=2\sqrt{\pi}i\mu(1-b\mu^2)a(\mu), \qquad \mu>0.
\eqno{(4.7)}
$$

Пользуясь формулами Сохоцкого для вспомогательной и дисперсионной функций,
от уравнения (4.4) приходим к краевому условию:
$$
\lambda^+(\mu)\bigg[N^+(\mu)-2U_0+A+
\dfrac{\eta_0(1-b\mu\eta_0)a_0}{\sqrt{\pi}(\eta_0-\mu)}\bigg]-
$$
$$
-\lambda^-(\mu)\bigg[N^-(\mu)-2U_0+A+
\dfrac{\eta_0(1-b\mu\eta_0)a_0}{\sqrt{\pi}(\eta_0-\mu)}\bigg]=0, \quad \mu>0.
\eqno{(4.8)}
$$

Рассмотрим соответствующую однородную краевую задачу Римана:
$$
X^+(\mu)=G(\mu)X^-(\mu),\qquad \mu>0,
\eqno{(4.9)}
$$
где коэффициент задачи $G(\tau)$ определен равенством (3.2).

Решение задачи Римана (4.9) проводится аналогично \cite{ALY-2} и дается
интегралом типа Коши:
$$
X(z)=\dfrac{1}{z^\varkappa}\exp V(z),
\eqno{(4.10)}
$$
где $\varkappa=\varkappa(G)$ -- индекс коэффициента задачи, введенный в п. 3, а
$V(z)$ понимается как интеграл типа Коши
$$
V(z)=\dfrac{1}{2\pi i}\int\limits_{0}^{\infty}
\dfrac{\ln G(\tau)-2\pi i \varkappa}{\tau-z}d\tau.
\eqno{(4.11)}
$$
Здесь $\ln G(\tau)=\ln |G(\tau)|+i \theta(\tau)$ -- главная ветвь логарифма,
фиксированная в нуле условием $\ln G(0)=0$, угол $\theta(\tau)=\arg G(\tau)$
-- главное значение аргумента, введенное равенством (3.3). Интеграл (4.11)
удобнее рассматривать в виде:
$$
V(z)=\dfrac{1}{\pi}\int\limits_{0}^{\infty}\dfrac{q(\tau)-\pi \varkappa}
{\tau-z},
$$
{где}
$$
q(\tau)=\dfrac{\theta(\tau)}{2}-\dfrac{i}{2}\ln|G(\tau)|,
$$
или в виде
$$
V(z)=\dfrac{1}{\pi}\int\limits_{0}^{\infty}\dfrac{\zeta(\tau)d\tau}
{\tau-z}, \qquad\varkappa=0,1,
$$
где
$$
\zeta(\tau)=q(\tau)-\pi \varkappa.
$$

Вернемся к решению неоднородной задачи (4.8), предварительно преобразовав
с помощью (4.9) ее к виду:
$$
X^+(\mu)\bigg[N^+(\mu)-2U_0+A+
\dfrac{\eta_0(1-b\mu\eta_0)a_0}{\sqrt{\pi}(\eta_0-\mu)}\bigg]-
$$
$$
-X^-(\mu)\bigg[N^-(\mu)-2U_0+A+
\dfrac{\eta_0(1-b\mu\eta_0)a_0}{\sqrt{\pi}(\eta_0-\mu)}\bigg]=0,
\quad \mu>0.
\eqno{(4.12)}
$$

Учитывая поведение всех входящих в краевое условие (4.12)
функций в комплексной плоскости и в бесконечно удаленной точке получаем
общее решение, из которого находим
$$
N(z)=2U_0-A+\dfrac{\eta_0(1-bz\eta_0)a_0}{\sqrt{\pi}(z-\eta_0)}+
\dfrac{1}{X(z)}\Big[C_0+\dfrac{C_1}{z-\eta_0}\Big],
\eqno{(4.13)}
$$
где $C_0, C_1$ -- произвольные постоянные, причем при $\varkappa=0\; C_1=0$,
а при $\varkappa=1\; C_0=0$, функция $X(z)$ определяется равенством (4.10).

Потребуем, чтобы общее решение (4.13) имело те же аналитические свойства,
что и вспомогательная функция (4.6). Рассмотрим случай $\varkappa=0$.
В этом случае  $C_0=-2U_0+A$.

Подставляя (4.13) в (4.7), находим коэффициент непрерывного спектра:
$$
a(\eta)=-\dfrac{2U_0-A}{2i\sqrt{\pi}\eta(1-b\eta^2)}
\Big[\dfrac{1}{X^+(\eta)}-\dfrac{1}{X^-(\eta)}\Big].
$$
Замечая, что
$$
\dfrac{1}{X^+(\eta)}-\dfrac{1}{X^-(\eta)}=-\dfrac{2i\sin\zeta(\eta)}{X(\eta)},
$$
с помощью (4.10) получаем следующее выражение для коэффициента
непрерывного спектра:
$$
a(\eta)=\dfrac{(2U_0-A)\sin \zeta(\eta)}{\sqrt{\pi}\eta(1-b\eta^2)X(\eta)}.
$$

Осталось найти выражение для $A$, входящее в последнее соотношение.
Подставляя это соотношение в (4.5), сначала находим величину $A$ из
полученного уравнения, а затем окончательно находим выражение
коэффициента непрерывного спектра:
$$
a(\eta)=\dfrac{2U_0\sin\zeta(\eta)}{\sqrt{\pi}(1+bJ_0)\eta(1-b\eta^2)X(\eta)},
\eqno{(4.14)}
$$
где
$$
J_0=\dfrac{1}{\pi}\int\limits_{0}^{\infty}\dfrac{\tau\sin\zeta(\tau)d\tau}
{(1-b\tau^2)X(\tau)}.
$$

В случае $\varkappa=1$ сначала устраняем полюс в точке $\eta_0$ условием
$$
C_1=-a_0\dfrac{\eta_0(1-b\eta_0^2)X(\eta_0)}{\sqrt{\pi}},
$$
затем, из условия $N(\infty)=0$ находим:
$$
C_1=-(2U_0-A)+\dfrac{a_0b\eta_0^2}{\sqrt{\pi}}.
$$

Далее, опуская длительные, но прямолинейные вычисления, на основании
(4.7), (4.13) и (4.5) находим сначала коэффициент дискретного спектра:
$$
a_0=\dfrac{2U_0\sqrt{\pi}}{b\eta_0^2+\eta_0(1-b\eta_0^2)X(\eta_0)(1-bJ_1)},
\eqno{(4.15)}
$$
где
$$
J_1=\dfrac{1}{\pi}\int\limits_{0}^{\infty}\dfrac{\tau \sin\zeta(\tau)d\tau}
{(1-b\tau^2)(\tau-\eta_0)X(\tau)},
$$
а затем и коэффициент непрерывного спектра
$$
a(\eta)=-\dfrac{C_1\sin\zeta(\eta)}{\sqrt{\pi}\eta(1-b\eta^2)
(\eta-\eta_0)X(\eta)},
\eqno{(4.16)}
$$
где
$$
C_1=-\dfrac{2U_0\eta_0(1-b\eta_0^2)X(\eta_0)}{b\eta_0^2+\eta_0(1-b\eta_0^2)
X(\eta_0)(1-bJ_1)}.
$$
На этом этапе доказательство разложения (4.1) (или (4.2)) закончено.

\begin{center}
\item{}\section{Функция распределения}
\end{center}

Рассмотрим функцию $h(x_1,\mu)$ для летящих
к стенке молекул непосредственно у стенки, т.е. при $x_1=0, \mu<0$.
Согласно (4.1) имеем:
$$
h(0,\mu)=\dfrac{1}{\sqrt{\pi}}\int\limits_{0}^{\infty}
\dfrac{\eta(1-b\mu\eta)}{\eta-\mu}a(\eta)d\eta, \quad\text{если}\quad\varkappa=0,
\eqno{(5.1)}
$$
и
$$
h(0,\mu)=a_0\dfrac{\eta_0(1-b\mu\eta_0)}{\sqrt{\pi}(\eta_0-\mu)}+
\dfrac{1}{\sqrt{\pi}}\int\limits_{0}^{\infty}
\dfrac{\eta(1-b\mu\eta)}{\eta-\mu}a(\eta)d\eta, \quad\text{если}\quad\varkappa=1.
\eqno{(5.2)}
$$
С помощью формулы (4.14) представим разложение (5.1) в явном виде:
$$
\dfrac{h(0,\mu)}{2U_0}=\dfrac{1}{(1+bJ_0)\pi}\int\limits_{0}^{\infty}
\dfrac{(1-b\mu\eta)\sin\zeta(\eta)d\eta}{(1-b\eta^2)X(\eta)(\eta-\mu)},
\eqno{(5.3)}
$$
а с помощью формул (4.15) и (4.16) в явном виде представим разложение (5.2):
$$
\dfrac{h(0,\mu)}{2U_0}=\dfrac{\eta_0(1-b\mu\eta_0)}
{[b\eta_0^2+\eta_0(1-b\eta_0^2)X(\eta_0)(1-bJ_1)](\eta_0-\mu)}+$$$$+
\dfrac{\eta_0(1-b\eta_0^2)X(\eta_0)}
{[b\eta_0^2+\eta_0(1-b\eta_0^2)X(\eta_0)(1-bJ_1)]}
\times $$$$ \times
\dfrac{1}{\pi}
\int\limits_{0}^{\infty}
\dfrac{(1-b\mu\eta)\sin\zeta(\eta)d\eta}
{(1-b\eta^2)X(\eta)(\eta-\eta_0)(\eta-\mu)}.
\eqno{(5.4)}
$$

Раccмотрим важный частный случай формул (5.3) и (5.4) при $b\to 0$,
т.е. когда рассматриваемое эллипсоидально--статис\-ти\-чес\-кое уравнение
переходит в БГК--уравнение. При $b=0$ из (5.3) и (5.4) соответственно имеем:
$$
\dfrac{h(0,\mu)}{2U_0}=\dfrac{1}{\pi}\int\limits_{0}^{\infty}
\dfrac{\sin\zeta(\eta)d\eta}{X(\eta)(\eta-\mu)},
\eqno{(5.5)}
$$
и
$$
\dfrac{h(0,\mu)}{2U_0}=\dfrac{1}
{X(\eta_0)(\eta_0-\mu)}+\dfrac{1}{\pi}\int\limits_{0}^{\infty}
\dfrac{\sin\zeta(\eta)d\eta}{X(\eta)(\eta-\eta_0)(\eta-\mu)}
\eqno{(5.6)}
$$ Формулы (5.5) и (5.6) в точности совпадают с соответствующими
формулами из нашей работы \cite{ALY-3}.

\begin{center}
\item{}\section{Скорость газа в полупространстве и непосредственно у стенки}
\end{center}

Согласно (1.3) безразмерная скорость газа в полупространстве
вычисляется  после подстановки (1.8) в (1.3) по формуле:
$$
U_y(x_1,t_1)=\dfrac{e^{-i\omega_1t_1}}{2\sqrt{\pi}}\int\limits_{-\infty}^{\infty}
e^{-\mu^2}h(x_1,\mu)d\mu.
\eqno{(6.1)}
$$

Заметим, что
$$
\int\limits_{-\infty}^{\infty}e^{-\mu^2}\Phi(\eta_0,\mu)d\mu=
\dfrac{\eta_0}{\sqrt{\pi}}\int\limits_{-\infty}^{\infty}\dfrac{e^{-\mu^2}d\mu}
{\eta_0-\mu}=1-i\omega_1=z_0,
$$
ибо $\eta_0$ -- нуль дисперсионной функции, т.е.
$$
\lambda(\eta_0)=-i\omega_1+1+\dfrac{\eta_0}{\sqrt{\pi}}
\int\limits_{-\infty}^{\infty}\dfrac{e^{-\mu^2}d\mu}{\mu-\eta_0}=0.
$$

Подставим разложение (4.1) в (6.1). Учитывая принятую выше
нормировку
$$
\int\limits_{-\infty}^{\infty}e^{-\mu^2}\Phi(\eta,\mu)d\mu=z_0,
$$
получаем следующее выражение для безразмерной скорости газа:
$$
U_y(x_1, t_1)=\dfrac{z_0e^{-i\omega_1t_1}}{2\sqrt{\pi}}
\Big[a_0e^{-x_1z_0/\eta_0}+\int\limits_{0}^{\infty}e^{-x_1z_0/\eta}
a(\eta)d\eta\Big].
\eqno{(6.2)}
$$

Пусть $\varkappa=0$. Подставим в (6.2) выражение (4.14) для
коэффициента непрерывного спектра. Тогда выражение для
размерной скорости газа имеет вид:
$$
\dfrac{u_y(x_1,t_1)}{u_0}=\dfrac{z_0e^{-i\omega_1t_1}}{(1+bJ_0)\pi}
\int\limits_{0}^{\infty}e^{-x_1z_0/\eta}\dfrac{\sin\zeta(\eta)d\eta}{\eta(1-b\eta^2)
X(\eta)}.
\eqno{(6.3)}
$$

Формула (6.3) служит для вычисления скорости газа в диапазоне
частот $\omega_1>\omega_1^*$.

Пусть $\varkappa=1$. Подставим в (6.2) коэффициенты непрерывного
спектра (4.16) и дискретного спектра (4.15). В результате для
размерной скорости газа получаем:
$$
\dfrac{u_y(x_1,t_1)}{u_0}=
\dfrac{z_0e^{-i\omega_1t_1}}{b\eta_0^2+y_0}\Bigg[e^{-\frac{x_1z_0}{\eta}}+
{\eta_0(1-b\eta_0)X(\eta_0)}\times $$$$ \times
\dfrac{1}{\pi}\int\limits_{0}^{\infty}
\dfrac{e^{-\frac{x_1z_0}{\eta}}\sin\zeta(\eta)d\eta}{\eta(1-b\eta^2)X(\eta_0)(\eta-\eta_0)}
\Bigg].
\eqno{(6.4)}
$$

Здесь
$$
y_0=y_0(\eta_0)=\eta_0(1-b\eta_0^2)X(\eta_0)(1-bJ_1).
$$

Формула (6.4) служит для вычисления скорости газа в диапазоне
частот $0\leqslant \omega_1 <\omega_1^*$.

Непосредственно у стенки из формул (6.3) и (6.4) при $x_1=0$
получаем значения скорости газа.  В случае $\varkappa=0$ и
$\varkappa=1$ соответственно имеем:
$$
\dfrac{u_y(0,t_1)}{u_0}=\dfrac{z_0e^{-i\omega_1t_1}}{(1+bJ_0)\pi}
\int\limits_{0}^{\infty}\dfrac{\sin\zeta(\eta)d\eta}{\eta(1-b\eta^2)
X(\eta)}
$$
и
$$
\dfrac{u_y(0,t_1)}{u_0}=\dfrac{z_0e^{-i\omega_1t_1}}{b\eta_0^2+y_0}
\Bigg[1+\eta_0(1-b\eta_0^2)X(\eta_0)\times $$$$ \times
\dfrac{1}{\pi}
\int\limits_{0}^{\infty}
\dfrac{\sin\zeta(\eta)d\eta}{\eta(1-b\eta^2)X(\eta_0)(\eta-\eta_0)}\Bigg].
\eqno{(6.5)}
$$

\begin{center}
\item{}\section{Гидродинамический характер решения}
\end{center}

Покажем, что при малых значениях $\omega_1$ скорость газа (6.5)
переходит в гидродинамическое выражение для скорости сплошной
среды, приведенное в \cite{LandauG}
$$
v=u_0e^{-x/\delta}e^{i(x/\delta-\omega t)}.
\eqno{(7.1)}
$$

Здесь $\delta=\sqrt{3\nu_{k.v.}/\omega}$, $\nu_{k.v.}$ --
кинематическая вязкость газа.

Формула (7.1) выведена для случая сплошной среды, когда
ограничивающая среду плоскость совершает гармонические колебания
по закону $u_s(t)=u_0e^{-i\omega t}$.

Заметим, что при $\omega_1\to 0$ нуль дисперсионной функции
$\eta_0^{(0)}=(1+i)/\sqrt{6\omega_1}\to \infty$,  а величина
$b\eta_0^2$ яляется ограниченной. Следовательно, интеграл по
непрерывному спектру в (6.5) исчезает и мы получаем:
$$
u_y(x_1,t)=\dfrac{u_0e^{-i(\omega t-x_1/\eta_0)}}{b\eta_0^2+
\eta_0(1-b\eta_0^2)X(\eta_0)(1-bJ_1)}.
$$

Заметим теперь, что при $\eta_0^{(0)}\to \infty$ $\eta_0 X(\eta_0)\to
1$, а $J_1\to 0$. Следовательно, из предыдущего выражения
вытекает, что
$$
u_y(x,t)=u_0e^{-i(\omega t-x\sqrt{\beta}/\tau \eta_0)}.
$$

Найдем величину $\sqrt{\beta}/\tau \eta_0$. Нетрудно проверить,
что для эллипсоидально -- статистического уравнения
кинематическая вязкость равна: $\nu_{k.v.}=\tau/3\beta$.
Следовательно,
$$
\dfrac{\sqrt{\beta}}{\tau \eta_0}=\dfrac{\sqrt{\beta}}{\tau}
\dfrac{\sqrt{6\omega \tau}}{1+i}=\sqrt{\dfrac{\beta}{\tau}}
\dfrac{\sqrt{6}}{2}(1-i)=\dfrac{1-i}{\sqrt{3\nu_{k.v.}/\omega}}=
\dfrac{1-i}{\delta}.
$$

Это означает, что последняя формула для скорости газа в точности
переходит в формулу (7.1) при $\omega_1\to 0$.

\begin{center}\bf
\item{}\section{Сила трения, действующая на колеблющуюся границу}
\end{center}

Компонента тензора вязких напряжений, приходящаяся на единицу площади
колеблющейся границы, вычисляется по формуле
$$
\sigma_{xy}(t)=m\int v_xv_yf(t,0,\mathbf{v})d^3v.
\eqno{(8.1)}
$$
Согласно \cite{LandauG} сила трения (приходящаяся на единицу площади),
действующая со стороны газа на пластину, равна
$
F_s(t)=-\sigma_{xy}(0,t).
$
Поэтому  согласно (8.1)
$$
F_s(t_1)=-e^{-i \omega_1t_1}\dfrac{p}{\sqrt{\pi}}
\int\limits_{-\infty}^{\infty}e^{-\mu^2}\mu h(0,\mu)d\mu, \quad p=nkT.
\eqno{(8.2)}
$$

Подставим в (8.2) разложение (4.1). Получаем, что
$$
F_s(t_1)=-i\omega_1e^{-i\omega_1t_1}\dfrac{p}{\sqrt{\pi}}\Big[a_0\eta_0+
\int\limits_{0}^{\infty}\eta a(\eta)d\eta\Big].
\eqno{(8.3)}
$$

Из формулы (8.3) в случае нулевого индекса ($\varkappa=0$) мы получаем
следующее выражение силы трения:
$$
F_s(t_1)=-i\omega_1e^{-i\omega_1t_1}\dfrac{2U_0p}{(1+bJ_0)\pi}
\int\limits_{0}^{\infty}\dfrac{\sin \zeta(\eta)d\eta}
{(1-b\eta^2)X(\eta)}.
$$

В случае единичного индекса ($\varkappa=1$) мы получаем следующее
выражение силы трения:
$$
F_s(t_1)=-i\omega_1e^{-i\omega_1t_1}\dfrac{2U_0p\eta_0}{(b\eta_0^2+y_0)}
\Bigg[1+(1-b\eta_0^2)X(\eta_0)\times $$$$ \times \dfrac{1}{\pi}
\int\limits_{0}^{\infty}\dfrac{\sin \zeta(\eta)d\eta}
{(1-b\eta^2)(\eta-\eta_0)X(\eta)}\Bigg].
$$
\begin{center}
\item{}\section{Заключение}
\end{center}

В настоящей работе сформулирована и решена аналитически вторая задача
Стокса --- задача о поведении
разреженного газа, занимающего полупространство над стенкой, совершающей
гармонические колебания. Рассматриваются диффузные граничные условия.
Используется линеаризованное эллипсоидально--статистическое
уравнение с параметром  $a=-2(1/\Pr-1)$, зависящем от числа Прандтдя $\Pr$.
На основе аналитического решения построена функция распределения
и найдена массовая скорость
разреженного газа в полупространстве.
Выявлен гидродинамический характер решения при малых значениях
частоты колебаний ограничивающей газ плоскости. Найдена сила трения,
действующая со стороны газа на колеблющуюся пластину.

\end{document}